%% file: main.tex
\documentclass{JFM-FLM_Au}
\usepackage{siunitx}
\usepackage{booktabs}
\usepackage{longtable}
\usepackage{array}
\usepackage[commandnameprefix=always,final]{changes}

\lefttitle{Kang Z.-C. et al.}
\righttitle{Journal of Fluid Mechanics}

\title{Multiple critical Froude numbers for the centrifugal effects on heat transport in rotating Rayleigh-B\'{e}nard convection}

\author{Zhi-Cong Kang\aff{1,2}\thanks{ Z.C.K and G.Y.D contributed equally to this work.}, Guang-Yu Ding\aff{3,2}$^{\dagger}$, Lu Zhang\aff{2}, \and Ke-Qing Xia\aff{2,4}}

\affiliation{
\aff{1}School of Aeronautics, Institute of Extreme Mechanics, Northwestern Polytechnical University, Xi'an 710072, PR China
\aff{2}Center for Complex Flows and Soft Matter Research, Department of Mechanics and Aerospace Engineering, Southern University of Science and Technology, Shenzhen 518055, PR China
\aff{3}Research Institute of Intelligent Complex Systems, Fudan University, Shanghai 200433, PR China
\aff{4}Department of Physics, Southern University of Science and Technology, Shenzhen 518055, PR China
}

\corresau{Ke-Qing Xia, \email{xiakq@sustech.edu.cn}; Guang-Yu Ding, \email{dinggy@fudan.edu.cn}}

\begin{document}
\maketitle

\begin{abstract}
The influence of centrifugal effects in rotating Rayleigh-B\'{e}nard convection is investigated using direct numerical simulations. It is found that the Nusselt number decreases beyond a critical Froude number, \chdeleted{namely} $Fr_c^*$. 
This critical value depends on both Rayleigh number $Ra$ and the aspect ratio $\Gamma$, following the scalings $Fr^*_c = 4.79 \times 10^{-2} \Gamma^{1.1\pm0.4}$ at fixed $Ra=8.71\times10^8$, and $Fr^*_c = 2.54 \times 10^{-13} Ra^{1.27\pm0.08}$ at fixed $\Gamma=1$. 
We interpret $Fr_c^*$ as \chdeleted{the critical Froude number for} the onset of centrifugal effects within the thermal boundary layers. This interpretation is supported by the thickening of the boundary layers and the drop in planar heat flux. 
We compare $Fr_c^*$ with \chdeleted{the} two previously proposed critical Froude numbers: $Fr_{Hu}=3.79\times10^{-7}Ra^{0.53}$, introduced by  \citeauthor{hu_jfm_2022} (\textit{J. Fluid Mech.}, vol. 938, 2022, R1), based on \chdeleted{the measurements of} bulk temperature anomaly; and $Fr_{Horn}=0.5\Gamma^1$, proposed by \citeauthor{horn_prl_2018} (\textit{Phys. Rev. Lett.}, vol. 120, 2018, 204502), based on a global force-balance argument. 
We confirm that $Fr_{Hu}$ marks the onset of centrifugal effects in the bulk, \chdeleted{as} evidenced by both local heat flux and radial vortex motion. 
For $Fr_{Hu}<Fr<Fr_c^*$, centrifugal effects have little impact on \chdeleted{the} global heat transfer, indicating that they mainly redistribute heat within the bulk without significantly altering boundary-layer dynamics. 
Meanwhile, $Fr_c^*$ exhibits a scaling with $\Gamma$ similar to that of $Fr_{Horn}$, suggesting a close connection between the global force balance and the onset of centrifugal effects within the thermal boundary layers. 
\chadded{
	Overall, our findings indicate that centrifugal force affects the bulk and the top and bottom boundary layers differently in rotating Rayleigh-B\'{e}nard convection.  
	A larger centrifugal force is required to modify the boundary layer properties. }
\end{abstract}



\input{sec_introduction}
\input{sec_numerical_method}
\input{sec_result_global_Nu}
\input{sec_result_bulk_onset}
\input{sec_result_BL_onset}
\input{sec_conclusion}

\begin{bmhead}[Acknowledgements.]
We gratefully acknowledge H.-D. Xi, F. Xu, J. Dong for helpful discussions.
\end{bmhead}
\begin{bmhead}[Funding.]
We gratefully acknowledge the support of this work from the National Natural Science Foundation of China (NSFC) (Grant Nos. 12232010, 12302282, 12202173, 12572251, 12595300, 12595302) and the Center for Computational Science and Engineering of Southern University of Science and Technology.
\end{bmhead}
\begin{bmhead}[Declaration of interests.]
The authors report no conflict of interest.
\end{bmhead}
\appendix
\input{sec_appendix}
\bibliographystyle{jfm}
\bibliography{reference}

\end{document}

%% file: sec_introduction.tex
\section{Introduction}
\label{sec:introduction}
Rotating convection widely exists in geophysical and astrophysical systems \citep{king_n_2009,guervilly_n_2019}. 
Rotational constraints are key to the formation of distinct flow structures in different planets, such as the zonal jets and polar vortex crystals in Jupiter \citep{christensen_grl_2001,kaspi_n_2018,siegelman_pnas_2022}. 
Rotation also plays an important role in the formation and maintenance of planetary magnetic fields \citep{glatzmaier_n_1999}. 
Studies of rotating convection help advance our understanding of these astrophysical phenomena.
As a result, rotating convection has attracted considerable attention \citep{kunnen_jt_2021, ecke_arfm_2023, xia_ams_2025}. 
A canonical model for studying rotating convection is rotating Rayleigh-B\'{e}nard convection (RRBC). 
This system refers to a rotating \chreplaced{container}{ system}  consisting of a fluid layer heated from below and cooled from above. 
The strength of buoyancy forcing and rotational constraint are respectively characterised by the Rayleigh number $Ra$ and the Ekman number $Ek$:

\begin{equation}
    \label{equ:def_ra_ek}
    Ra\equiv\frac{\alpha g\Delta H^3}{\kappa\nu},\ Ek\equiv\frac{\nu}{2\Omega H^2}.
\end{equation}
Here $\alpha$ denotes the adiabatic thermal expansion coefficient, $g$ is the gravitational acceleration, $\Delta$ is the temperature difference between the heating and cooling plates, $H$ is the cell height, $\kappa$ is the thermal diffusivity, $\nu$ is the viscosity, and $\Omega$ is the rotation rate. 
The working fluid is characterised by the Prandtl number $Pr\equiv \nu/\kappa$. This system is mathematically well defined and experimentally accessible, and is therefore widely used in both numerical and experimental studies \citep{ecke_arfm_2023,xia_ams_2025}. In recent decades, many interesting features of this system have been discussed, such as the various flow structures \citep{sprague_jfm_2006,julien_gafd_2012,stellmach_prl_2014,plumley_jfm_2016,cheng_prf_2020}, vortex dynamics \citep{sakai_jfm_1997,noto_jfm_2019,chong_sa_2020,ding_nc_2021}, etc.

In most previous studies, the centrifugal force has been neglected. 
The strength of centrifugal effects can be quantified by the Froude number 

\begin{equation}
	\label{equ:fr_def}
	Fr\equiv\frac{\Omega^2 R}{g},
\end{equation}
where $R$ is the radius of the convection cell. 
Centrifugal effects are generally considered negligible for $Fr<0.05$ \citep{marques_jfm_2007}.
However, as the rotation rate increases, centrifugal effects become progressively important and have recently attracted growing attention.

There are two main influences arising from the centrifugal effects in RRBC.
First, it can suppress global heat transport. 
The heat transport efficiency is quantified by the Nusselt number

\begin{equation}
\label{equ:def_nu}
    Nu\equiv\frac{q}{k\Delta/H}=\frac{\langle u_zT-\kappa\partial T/\partial z\rangle}{\kappa \Delta/H},
\end{equation}
where $q$ is the vertical heat flux, $k$ is the thermal conductivity of the working fluid, $T$ is the temperature, and $\mathbf{u}$ is the velocity vector, the subscript $z$ refers to the vertical direction/axis  (see also the $\partial z$), and $\langle\cdot\rangle$ denotes the averaging over time and horizontal cross-section.
\citet{horn_prl_2018,horn_prf_2019} identified a regime in which \chadded{both} Coriolis and centrifugal buoyancy dominate (CC regime), leading to a reduction in heat transport.
They proposed a critical Froude number $Fr_{Horn}=\Gamma/2$ for this regime.
Here $\Gamma\equiv D/H$ is the aspect ratio. 
This critical Froude number is obtained from the balance between the centrifugal buoyancy time scale $\tau_{cb}$ and the free-fall time scale $\tau_{ff}$, and characterises a global transition of the system.

Second, centrifugal force breaks the horizontal translational symmetry of the system and induces a radial separation of hot and cold fluids.  
Recent experimental studies conducted by \citet{hu_jfm_2022} propose another critical Froude number based on temperature measurements, \chdeleted{which is} denoted \chreplaced{as}{ by} $Fr_{Hu}$ hereinafter. 
They found that hot and cold flows begin to separate by centrifugal force as $Fr$ reaches $Fr_{Hu}=3.79\times10^{-7}Ra^{0.53}$ \chadded{in a $\Gamma=1$ cell}. 
A different set of experiments using an off-center rotating convection apparatus provides additional evidence supporting this scaling relationship \citep{hu_prl_2021,hu_jfm_2023}.
Recently, a $\Gamma=2$ result \citep{hu_jfm_2026} shows good agreement with the previously $\Gamma=1$ result on $Fr_{Hu}$, and the authors propose a local force balance argument to explain it. This suggests that $Fr_{Hu}$ is independent of $\Gamma$.
\chadded{Furthermore,} experiments and numerical studies focusing on the motion of convective Taylor columns (CTCs) reveal that centrifugal force can give rise to the inward (outward) motion of the hot (cold) columns, and provide a theoretical model based on a modified Langevin equation for the dynamics of CTCs \citep{chong_sa_2020,ding_nc_2021,ding_jfm_2023_vortex}. 

In addition to the convective Taylor columns, centrifugal effects also strongly influence the flow near the lateral sidewalls, which contribute significantly to the convective heat flux in RRBC \citep{dewit_prf_2020,zhang_prl_2020,lu_prf_2021,ding_jfm_2023}. 
Hereinafter, we refer to this sidewall-attached flow as the boundary flow, noting that the term is used differently in the literature. 
This boundary flow typically exhibits an azimuthally organised structure, in which hot and cold fluids are arranged in an alternating pattern with a well-defined mode number.
Recent numerical simulations reveal that centrifugal force helps trigger the instability of the boundary flow and leads to a drop in both local and global heat transport \citep{ding_jfm_2023}. 
Additionally, the drop in $Nu$ observed in \citet{ding_jfm_2023} occurs at $Fr<Fr_{Horn}$, suggesting that $Fr_{Horn}$ acts as an upper bound on heat transfer. 
However, the detailed response of the boundary flow to increasing centrifugal effects remains poorly understood. 

It \chreplaced{is}{ remains} unclear how the two critical Froude numbers $Fr_{Hu}$ and $Fr_{Horn}$ are connected and how the heat transport of the system evolves from local to global centrifugal effects.
As rotating experiments have \chreplaced{expanded the}{ pushed}  parameter range\chdeleted{s} in recent years \citep{cheng_gafd_2018,cheng_prf_2020}, the influence of centrifugal force and the boundary flow has become \chreplaced{increasingly}{ progressively more} significant, yet remains far from clear. 
To advance our understanding of this issue, we conduct a series of numerical simulations of RRBC with varying $Fr$, focusing on the influence of centrifugal effects on flow dynamics and heat transport. 
The rest of this paper is organised as follows: In Section \ref{sec:numerical_method}, we briefly introduce the parameter range of this study, as well as the numerical methods.
In Section \ref{sec:results}, we \chreplaced{introduce}{ reveal} a new critical Froude number $Fr_c^*$ for the global heat transport, and compare it with $Fr_{Hu}$ and $Fr_{Horn}$.
In Section \ref{sec:conclusion}, a conclusion is provided. 

%% file: sec_numerical_method.tex
\section{Numerical method}\label{sec:numerical_method}
In this study, we perform direct numerical simulations (DNS) of rotating Rayleigh-B\'{e}nard convection in a cylindrical domain. 
No-slip boundary conditions are applied at all surfaces. 
The top and bottom plates are at constant temperatures $T_{cold}$ and $T_{hot}$, respectively, and the lateral boundaries are adiabatic. 
The characteristic length $l^*$, time $t^*$ and temperature $T^*$ of the system are given by the cell height $H$, free-fall time scale $\sqrt{H/(\alpha g\Delta)}$ and the imposed temperature difference $\Delta=T_{hot}-T_{cold}$, respectively. 
We apply the Oberbeck-Boussinesq approximation and use $l^*$, $t^*$ and $T^*$ for nondimensionalization, obtaining the dimensionless governing equations

\begin{equation}
    \label{equ:governing_equ_mom}
    \frac{\partial \boldsymbol{u}}{\partial t} + \boldsymbol{u} \cdot \nabla \boldsymbol{u} = -\nabla p + \sqrt{\frac{Pr}{Ra}} \nabla^2 \boldsymbol{u} - \frac{1}{Ek} \sqrt{\frac{Pr}{Ra}} \hat{e}_z \times \boldsymbol{u} + \theta \hat{e}_z - Fr \frac{2r}{\Gamma} \theta \hat{e}_r,
\end{equation}
\begin{equation}
    \frac{\partial \theta}{\partial t} + \boldsymbol{u} \cdot \nabla \theta = \frac{1}{\sqrt{Ra Pr}} \nabla^2 \theta.
\end{equation}
\begin{equation}
    \nabla \cdot \boldsymbol{u} = 0,
\end{equation}
where $\boldsymbol{u}$ is the velocity, $p$ is pressure, $\theta=(T-T_m)/\Delta$ is temperature deviation, and $T_m=(T_{hot}+T_{cold})/2$ is the dimensional mean temperature. 
In our simulations, we fix $Pr=4.38$, corresponding to water at about \SI{40}{\celsius}, and $Ek = 1.85 \times 10^{-6}$. 
To explore the influence of centrifugal force, we treat $Fr$ as the main variable, ranging from 0 to 1 in the current study. 
As discussed in Section \ref{sec:introduction}, the critical Froude numbers $Fr_{Hu}$ and $Fr_{Horn}$ exhibit distinct parameter dependences: \chreplaced{The former scales with $Ra$ and is inferred to be independent of $\Gamma$, whereas the latter depends only on $\Gamma$.}{The former scales with $Ra$, whereas the latter scales with $\Gamma$.}
To examine centrifugal effects from both perspectives, we conduct two sets of simulations, each with a different secondary variable in addition to $Fr$. 
In Set I, we examine the influence of aspect ratio. Three aspect ratios are considered: $\Gamma=0.5,\ 1,\ 2$. 
The Rayleigh number is fixed at $Ra = 8.71\times10^{8}$ in this set, which is above the onset of bulk convection but below the breakdown of the boundary flow, according to \citet{ding_jfm_2023}. 
In Set II, we examine the centrifugal effects under different buoyancy forcing strengths, and change $Ra$ from $3.40\times10^{8}$ to $1.30\times10^{9}$, with \chadded{a fixed} $\Gamma=1$ \chdeleted{fixed}. 
Set II covers the onset of bulk convection and the breakdown of the boundary flow.  
Phase diagrams of the two sets of simulations are presented in figure~\ref{fig:phase_paragram}.
A detailed summary of all cases is provided in table \ref{tab:data} in the appendix.

\begin{figure}
	\centerline{\includegraphics[width=1.0\textwidth]{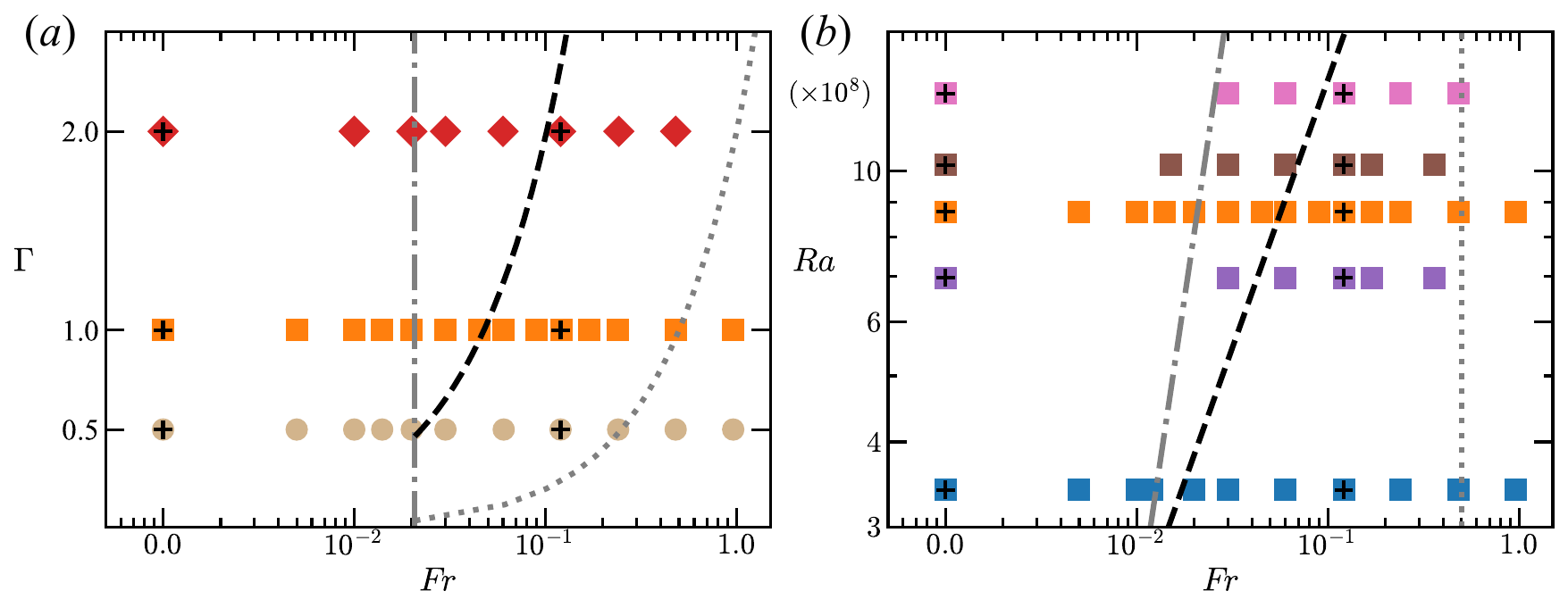}}
	\caption{Phase diagrams of the two sets of simulations: (a) Set I at fixed $Ra=8.71\times10^{8}$ and (b) Set II at fixed $\Gamma=1$. The cross symbol marks the parameter location reported by \citet{ding_jfm_2023}. The grey dash-dotted line and the dotted line indicate the critical Froude number $Fr_{Hu}$ \citep{hu_jfm_2022} and $Fr_{Horn}$ \citep{horn_prl_2018}, respectively, while the black dashed line denotes the critical Froude number $Fr_c^*$ proposed in the present study.}
	\label{fig:phase_paragram}
\end{figure}

The numerical method used in this study is the same as that in \citet{ding_jfm_2023}.
The governing equations are solved using a well-tested code called CUPS, which is developed based on a fourth-order finite volume method \citep{kaczorowski_jfm_2013,kaczorowski_jfm_2014,chong_jcp_2018}. 
As we set $Pr=4.38$ in this study, a multiple resolution scheme can be applied to enhance numerical efficiency. 
In this scheme, the velocity and temperature fields are solved on different meshes, allowing them to resolve the Kolmogorov and Batchelor length scales, respectively. 
The ratio between velocity and temperature grid spacing is denoted by $M_i$.  
Here, the subscript $i=r,\phi,z$ corresponds to the radial, azimuthal, and vertical directions, respectively.
To resolve the Ekman layer in our simulations, we set $M_z=1$ and $M_r=M_\phi=2$. 
Details of this scheme can be found in \citet{chong_jcp_2018}. All statistics (except for $\Gamma=2$) are averaged over at least 400 free-fall time units. 
For $\Gamma=2$, 200 free-fall time units are used because of the computational cost. To ensure the boundary layers are well resolved, there are at least 4 grid points within the Ekman boundary layer, and 9 within the boundary flows.

%% file: sec_result_global_Nu.tex
\section{Results}
\label{sec:results}
\begin{figure}
	\centerline{\includegraphics[width=1.0\textwidth]{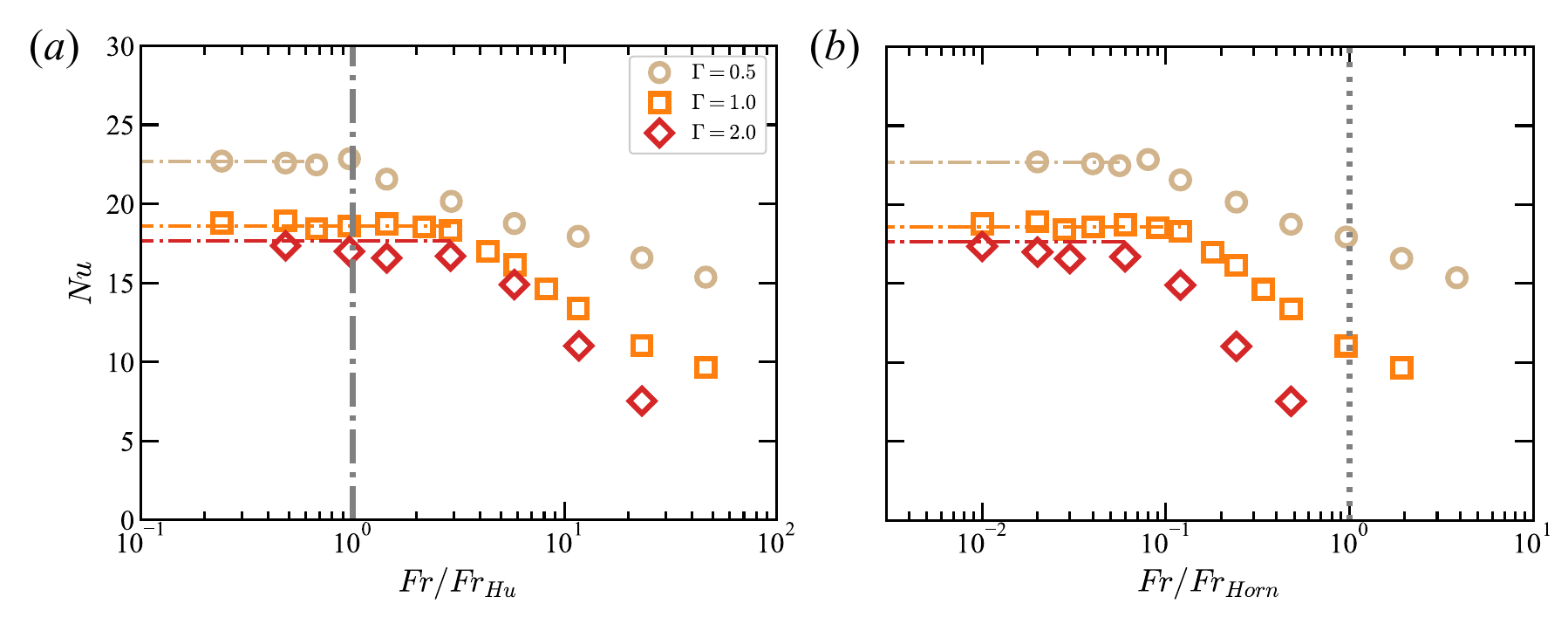}}
	\caption{Plot of $Nu$ as a function of (a) $Fr/Fr_{Hu}$ and (b) $Fr/Fr_{Horn}$. The vertical grey dash-dotted line in (a) indicates the critical Froude number determined by \citet{hu_jfm_2022}, while the vertical grey dotted line in (b) marks the critical Froude number reported by \citet{horn_prl_2018}. The horizontal dash-dotted lines represent the $Nu$ of the case without centrifugal force.}
	\label{fig:global_nu}
\end{figure}

To examine the influence of centrifugal force, we first consider the global heat transport efficiency $Nu$ for Set I.
We plot $Nu$ as a function of $Fr/Fr_{Hu}$ and $Fr/Fr_{Horn}$ in figure~\ref{fig:global_nu}. 
A clear reduction in $Nu$ is observed once $Fr$ exceeds a critical value, although this threshold is not explicitly indicated in figure~\ref{fig:global_nu}.
The quantitative definition of $Fr_c^*$ and the associated analysis are presented in Section \ref{subsec:bl_onset}. 
Qualitatively, it is evident that neither $Fr_{Hu}$ nor $Fr_{Horn}$ cannot account for $Fr_c^*$, according to figure~\ref{fig:global_nu}. 
The drop in $Nu$ occurs at values of $Fr$ clearly larger than $Fr_{Hu}$, except for $\Gamma=0.5$. 
Additionally, $Fr_{Hu}$ does not address the dependence on aspect ratio $\Gamma$, while figure~\ref{fig:global_nu}(a) suggests that $Fr_c^*$ increases with $\Gamma$. 
On the other hand, centrifugal effects begin to affect the global heat transport at much smaller $Fr$  than $Fr_{Horn}$ for the current setup, as shown in figure~\ref{fig:global_nu}(b). 
Since the thermal boundary layer controls the global heat transport in the present parameter range, $Fr^*_c$ can be considered as the onset of centrifugal effects within the boundary layer. 
We see that $Fr_c^*$ deviates from both the bulk temperature anomaly $Fr_{Hu}$ and the global force-balance relationship $Fr_{Horn}$. 
This indicates that the centrifugal effects do not act uniformly throughout the system but vary from region to region, with different parts of the flow responding in distinct ways.
To capture this spatial variation, decomposing the local heat flux may provide deeper insight.

%% file: sec_result_bulk_onset.tex
\subsection{Bulk centrifugal effects}
\label{subsec:bulk_onset}

\subsubsection{Horizontal heat exchange}

\begin{figure}
	\centerline{\includegraphics[width=1.0\textwidth]{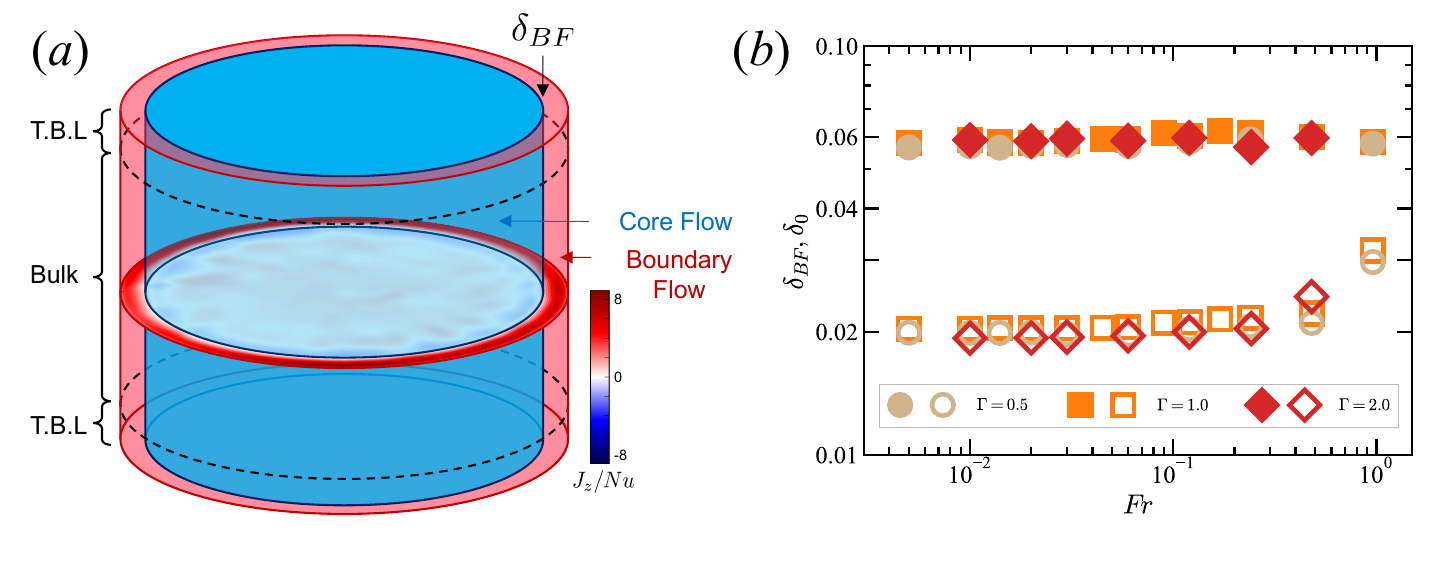}}
	\caption{(a) Sketch of the different regions in the system. The blue and red shaded regions denote the core flow (CF) and boundary flow (BF) defined in the radial direction. The interface between these two regions is indicated by $\delta_{BF}$, as shown by the black arrow.  The dashed curves near the top and bottom plates denote the thermal boundary layers (T.B.L.), separating the bulk from the thermal boundary layers in the vertical direction. The contour corresponds to the ratio $J_z/Nu$ at the mid-height, where $J_z=\sqrt{RaPr} u_z\theta$ is the convective heat flux. (b) Widths of the boundary flow thickness $\delta_{BF}$ (solid symbols) and outer layer thickness $\delta_{0} $ (open symbols) as a function of $Fr$.}
	\label{fig:bf_thickness}
\end{figure}

In addition to the conventional division into the top and bottom boundary layers and the bulk region in the vertical direction, we further distinguish the core flow (CF) and boundary flow (BF) in the radial direction.
Figure~\ref{fig:bf_thickness}(a) illustrates this decomposition and shows the heat flux ratio $J_z/Nu$ at the mid-height plane.
As seen in figure~\ref{fig:bf_thickness}(a), the convective heat flux at the boundary flow is much larger than that in the core flow, indicating strong spatial inhomogeneity in heat transport.
Therefore, it is necessary to examine the local heat flux in these two regions separately. 

Following the method proposed by \citet{ding_jfm_2023}, we first decompose the core and boundary flows. 
These two regions are identified according to the root-mean-square profile of the azimuthal velocity $\langle u_{\phi}^2\rangle_{t,z,\phi}^{1/2}(r)$. 
As shown in \citet{ding_jfm_2023}, the boundary flow can be properly described as a two-layer structure. 
By fitting the inner peak (closer to the axis) with parabolic functions, the width of the boundary flow $\delta_{BF}$ can be determined according to the intersection between the core profile and the fitted parabolic function. 
The width of the outer boundary flow $\delta_0$ is then determined by the zero-crossing of the parabolic fitting for the outer peak. 
For more details, one can refer to \citet{ding_jfm_2023}.

Figure~\ref{fig:bf_thickness}(b) presents $\delta_{BF}$ and $\delta_0$ as a function of $Fr$, where open symbols denote the outer layer thickness $\delta_0$ and solid symbols the boundary flow thickness $\delta_{BF}$. 
For the width of the outer layer $\delta_0$, we observe an increase when $Fr > 0.1$. 
One possible explanation is the deformation of the boundary flow under strong centrifugal effects, but a detailed analysis is beyond the scope of the present study.
On the other hand, the width of boundary flow $\delta_{BF}$ exhibits no clear dependence on $Fr$, which is consistent with \citet{ding_jfm_2023} but over a much broader range of $Fr$. 
It should be noted that the present analysis only considers the geometric thickness of the boundary flow, while its strength is not explicitly quantified. 
Therefore, the observed $Fr$-independence applies primarily to its spatial extent.
Within the parameter range explored here, $\delta_{BF}$ also shows no clear dependence on the aspect ratio $\Gamma$.
Taken together, these results indicate that the boundary-flow thickness is largely insensitive to both centrifugal effects and the aspect ratio, although a more complete assessment would require dynamical measures of the flow intensity.

\begin{figure}
	\centerline{\includegraphics[width=1.0\textwidth]{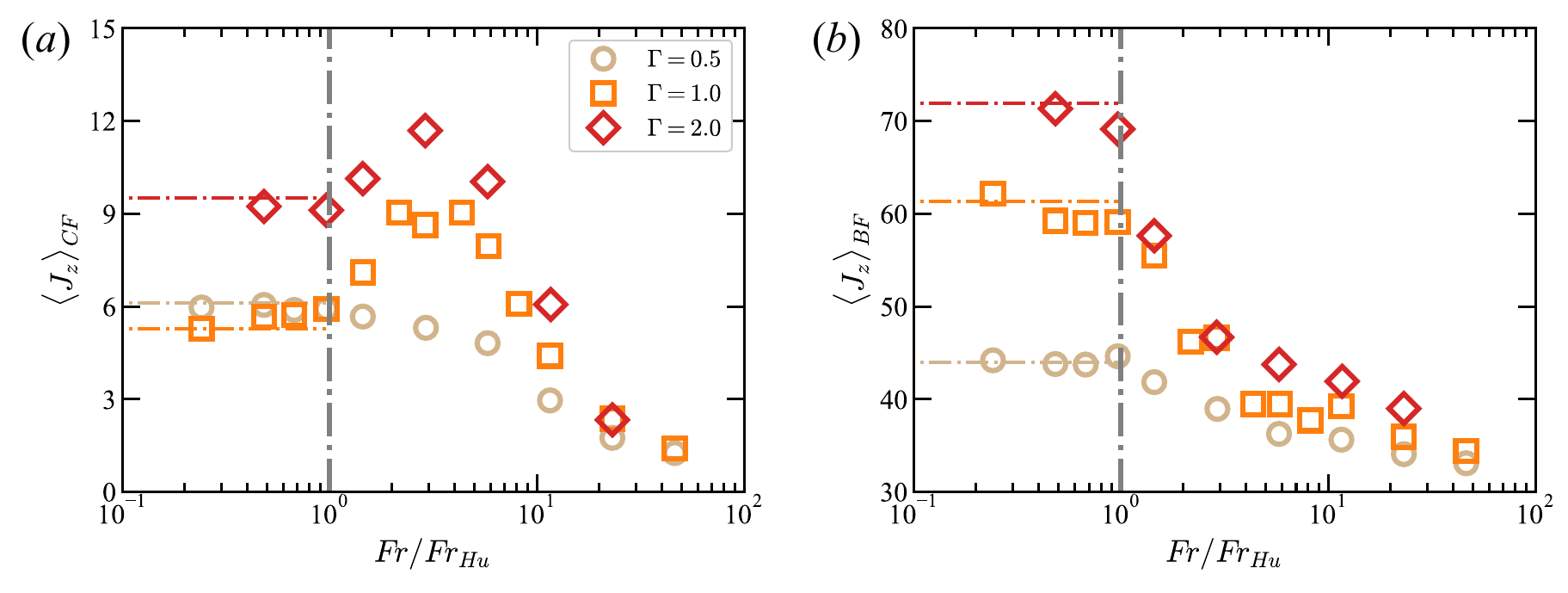}}
	\caption{ Plot of vertical convective heat flux (a) for the core flow $\langle J_z \rangle_{CF}$ and (b) for the boundary flow $\langle J_z \rangle_{BF}$  as a function of $Fr/Fr_{Hu}$. The horizontal dash-dotted lines and the vertical grey dash-dotted line are the same as those in Figure \ref{fig:global_nu}.}
	\label{fig:decomp_nu}
\end{figure}

Given this geometric characterisation, we now turn to the heat transport properties of the two regions. Specifically, we examine the volume-averaged convective heat flux for the core $\langle J_z\rangle_{CF}$ and boundary flows $\langle J_z\rangle_{BF}$, defined as 

\begin{equation}
	\langle J_z\rangle_{CF} \equiv \sqrt{RaPr}\left< u_z \theta \right>_{t,z,\phi,r\leq r_{BF}},
\end{equation}
\begin{equation}
	\langle J_z\rangle_{BF}   \equiv \sqrt{RaPr} \left<u_z \theta \right>_{t,z,\phi,r\geq r_{BF}}.
\end{equation}
Here $r_{BF}=R-\delta_{BF}$ is the radius at the CF--BF interface.
\chadded{
	The relationship between $\langle J_z\rangle_*$ and $Nu$ is:
	\begin{equation}
		Nu = f_{CF}\,\langle J_z\rangle_{CF} + (1-f_{CF})\,\langle J_z\rangle_{BF} + 1,
	\end{equation}
	where $f_{CF}=r_{BF}^2/R^2$ is the volume fraction of the core flow.}

Figure~\ref{fig:decomp_nu}(a) and \ref{fig:decomp_nu}(b) show $\langle J_z\rangle_{CF}$ and $\langle J_z\rangle_{BF}$ as a function of $Fr/Fr_{Hu}$, respectively. 
Interestingly, we observe significant changes in magnitude for both $\langle J_z\rangle_{CF}$ and $\langle J_z\rangle_{BF}$, once $Fr$ exceeds $Fr_{Hu}$ (i.e. $Fr/Fr_{Hu}>1$). 
For $\Gamma=1$ and $2$, $\langle J_z\rangle_{CF}$ first increases 
when $Fr \geq Fr_{Hu}$, and then decrease after reaching a maximum. 
For $\Gamma=0.5$, as $Fr\geq Fr_{Hu}$, $\langle J_z\rangle_{CF}$ 
begins to drop (with no initial increase). 
On the other hand, \chadded{for all values of $\Gamma$,} $\langle J_z\rangle_{BF}$ decreases once $Fr/Fr_{Hu}\geq 1$, and no enhancement is observed. 

By comparing figures \ref{fig:global_nu} and \ref{fig:decomp_nu}, one can find that a change in local heat flux does not necessarily lead to a similar change in global Nusselt number. 
The local convective heat flux $\langle J_z\rangle_{CF}$ and $\langle J_z\rangle_{BF}$ are averaged over control volumes, as illustrated by the blue- and red-shaded regions in figure~\ref{fig:bf_thickness}(a). 
These two regions are not dynamically isolated, and radial heat exchange occurs between them. 
As a result, the convective heat fluxes $\langle J_z\rangle_{CF}$ and $\langle J_z\rangle_{BF}$ should not be interpreted as entirely independent contributions. 
For $Fr>Fr_{Hu}$, the local centrifugal effects emerge and drive fluid motion in the radial direction, which leads to heat exchange between the core and boundary flows. 
In other words, the heat exchange between CF and BF suggested by figure~\ref{fig:decomp_nu} characterises the emergence of centrifugal effects in the bulk, thus the change in $\langle J_z\rangle_{CF}$ and $\langle J_z\rangle_{BF}$ is consistent with $Fr_{Hu}$.
On the other hand, the thermal boundary layer remains unaffected until $Fr_c^*$ is reached.
For $Fr>Fr_c^*$, the thermal boundary layer is modified by the centrifugal force, so that the global heat transport begins to decrease.

\subsubsection{Radial vortex motion}
\label{subsubsec:radial_motion}
In this section, we continue to discuss the centrifugal effects in the bulk. Radial vortex motion provides a direct indicator of centrifugal effects in the bulk flow. 
In the current study, vortices are identified by applying the $Q$-criterion \citep{hunt_stunsd2p1sp_1988} to the data at the height $z/H=0.2$. 
The quantity $Q$ is defined by

\begin{equation}
	Q \equiv \frac{1}{2}\left(\lVert \omega \rVert^2-\lVert S \rVert^2\right),
\end{equation}
where $\omega$ is the vorticity tensor and $S$ is the symmetric strain tensor. A vortex is identified as a connected region satisfying $Q>Q_{std}$, where $*_{std}$ denotes the standard deviation. The vortex center $\mathbf{r}_c$ is then defined as the weighted average location vector of the $Q$ distribution within a vortex 

\begin{equation}
	\mathbf{r}_c\equiv \frac{\int_{vortex}Q\mathbf{r}dA}{\int_{vortex} QdA}.
\end{equation}

\begin{figure}
	\centerline{\includegraphics[width=1.0\textwidth]{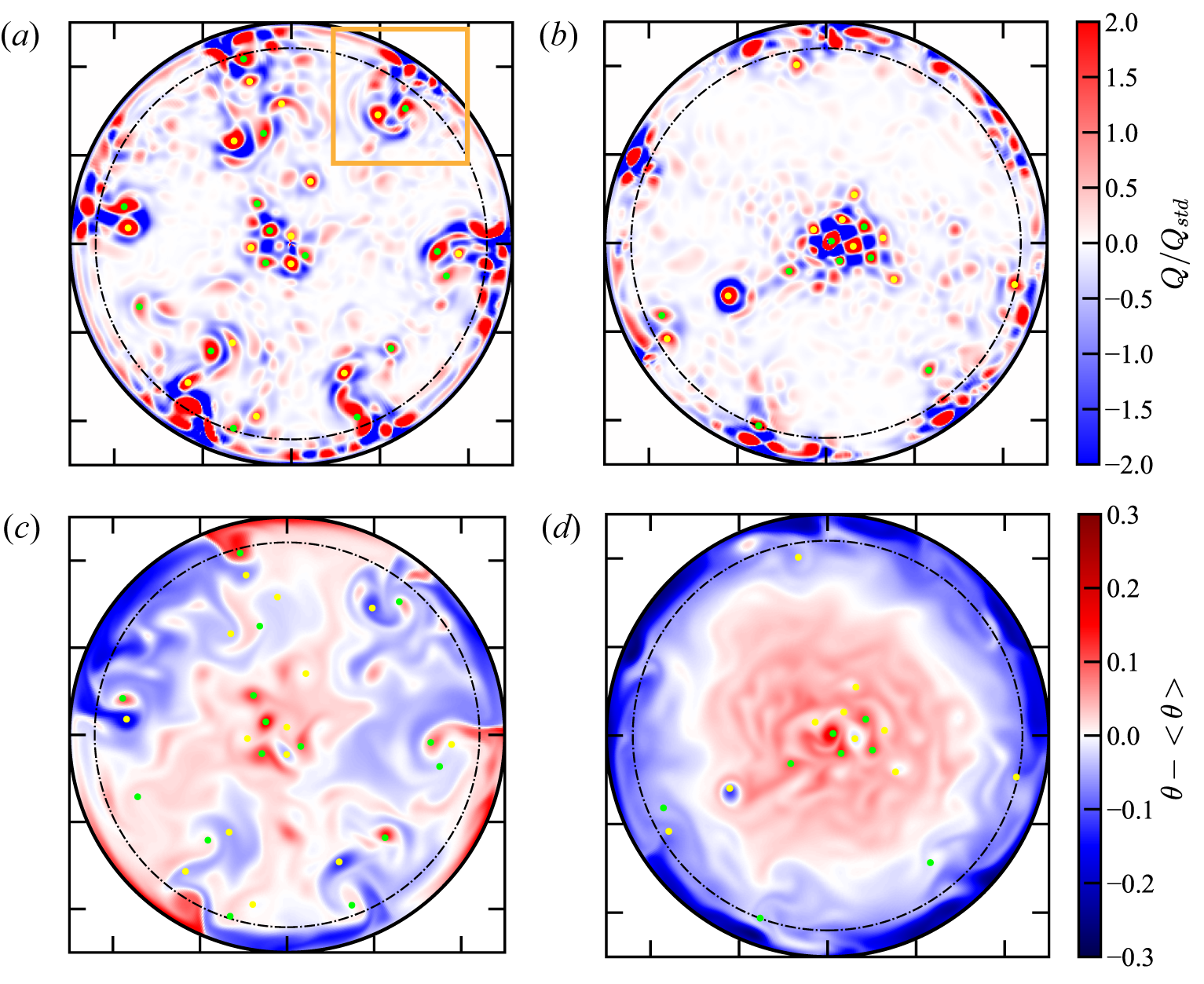}}
	\caption{(a,b) Instantaneous $Q/Q_{std}$ and (c,d) the corresponding temperature difference $\theta-\langle \theta \rangle$ at the height $z/H=0.2$ for (a,c) $Fr=0$ and (b,d) $Fr=0.48$, respectively. $Q_{std}$ is the standard deviation of $Q$ and $\langle \theta \rangle$ denotes the mean of $\theta$ at the same height. All plots are for $Ra=8.71\times10^{8},\Gamma = 1$. The green and yellow dots represent the centres of cyclones and anticyclones. The black dash-dotted circles correspond to $\delta_{BF}$. The area indicated by the orange square in (a) is enlarged in the following figure.}
	\label{fig:vortex_T}
\end{figure}

\begin{figure}
	\centerline{\includegraphics[width=1.0\textwidth]{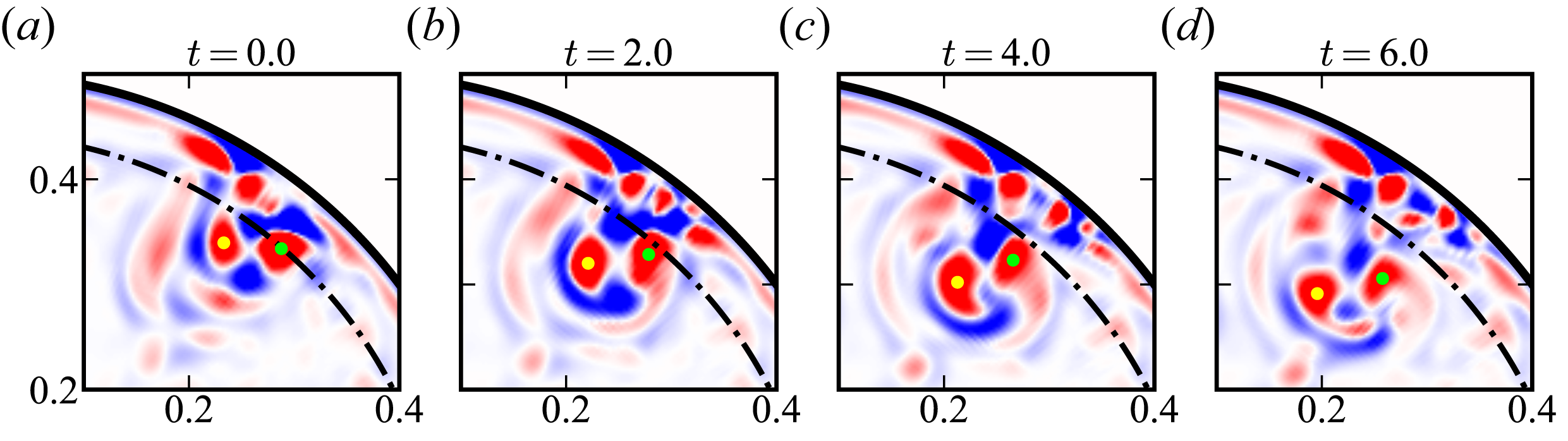}}
	\caption{Temporal evolution of the $Q/Q_{std}$ within the orange square in figure~\ref{fig:vortex_T}(a). Panels (a-d) correspond to $t=0.0,2.0,4.0,$ and $6.0$ in free-fall time units, respectively. \chadded{The snapshot at $t=6.0$ in panel (d) is exactly the one shown in figure~\ref{fig:vortex_T}(a).} The \chadded{meanings of} green dot, the yellow dot, and the black dash-dotted circles are \chreplaced{the same as}{ consistent with} those in figure~\ref{fig:vortex_T}.}
	\label{fig:vortex_emit}
\end{figure}

Figure~\ref{fig:vortex_T}(a,b) presents the normalized $Q$-fields $Q/Q_{std}$ for $Fr=0$ and $0.48$ \chadded{, respectively}, while panels (c,d) show the corresponding temperature fields. 
The green and yellow dots indicate the centers \chadded{$\mathbf{r}_c$} of the cyclones and anticyclones, respectively. 
At this height, cyclones correspond to the hot vortices and the anticyclones to the cold ones. 
The dash-dotted circles denote the edge of the boundary flow. From figure~\ref{fig:vortex_T}(c), it is evident that the boundary flow is azimuthally periodic with a mode number $N=3$. 
At the interfaces between the hot and cold regions, vortices are emitted to the core flow, as shown in figure~\ref{fig:vortex_T}(a). 
Figure~\ref{fig:vortex_emit} provides the temporal evolution of the two vortices located in the orange square in figure~\ref{fig:vortex_T}(a). 
It \chreplaced{shows}{ is observed} that vortices are generated near the sidewall and subsequently emitted to the core flow region.
This vortex emission from the boundary flow becomes less frequent when $Fr$ increases to 0.48, as shown in figures~\ref{fig:vortex_T}(b) and \ref{fig:vortex_T}(d). 
As $Fr$ increases, the centrifugal force creates a radial separation of hot and cold fluids, and meanwhile reduces temperature differences along the azimuthal direction within the boundary flow.
Such enhanced azimuthal uniformity may explain the observed suppression of vortex emission.
In this regime, cyclones are clustered with anticyclones, the latter of which move toward the sidewalls under the influence of the centrifugal force. 
Additionally, the mode number of the boundary flow becomes subtle in this case. Nevertheless, the boundary flow clearly persists even at such a large Froude number. It is evident that the boundary flow region can be identified from both the $Q$-field and the temperature distribution.

\begin{figure}
	\centerline{\includegraphics[width=0.9\textwidth]{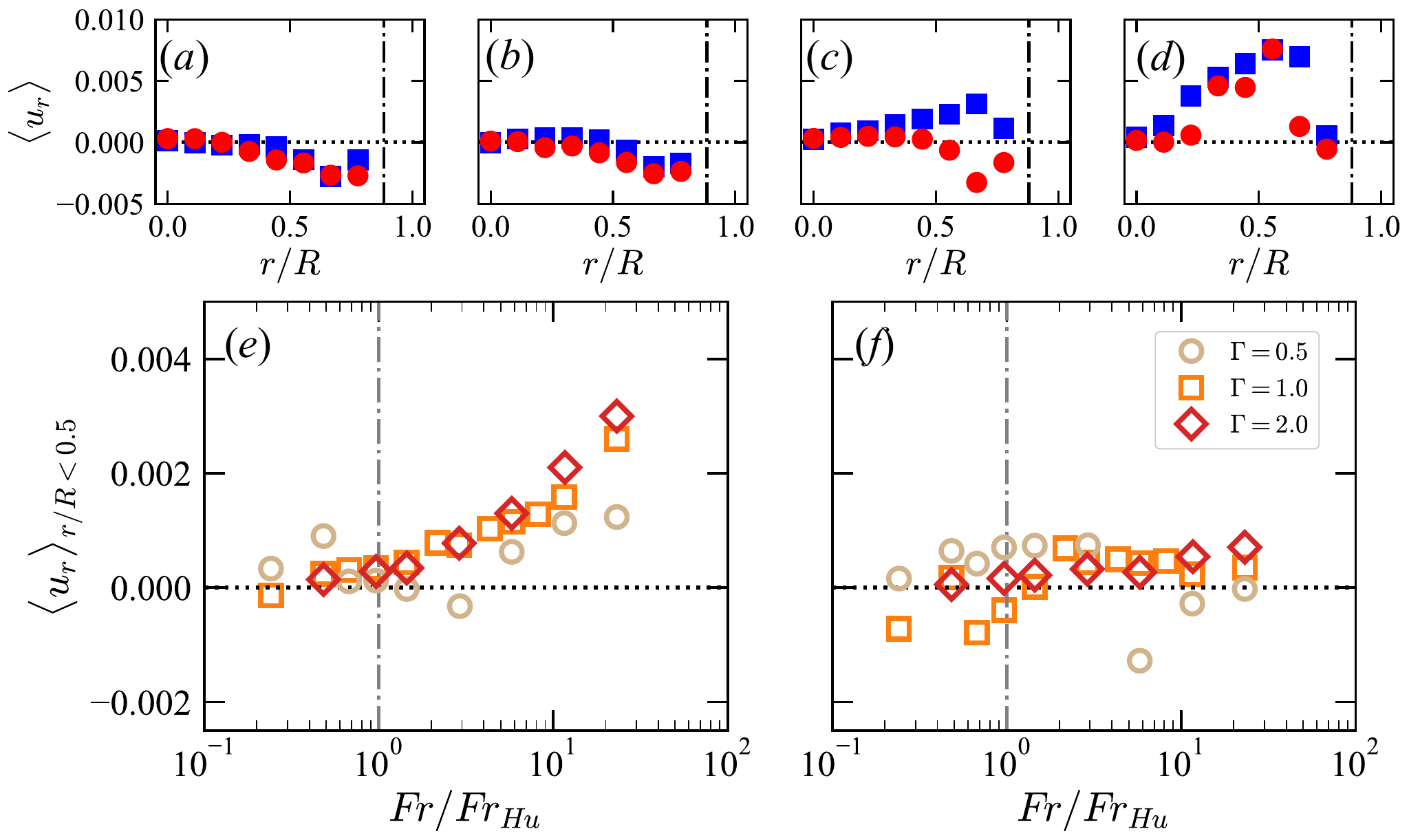}}
	\caption{(a-d) Radial profiles \chreplaced{$\langle u_r\rangle$}{ $\overline{u}_r$} of cyclones (red circles) and anticyclones (blue squares) for $Fr = 0, 0.02, 0.12,$ and $0.48$, respectively. All plots are for $Ra=8.71\times10^{8}, \Gamma=1.0$ \chadded{, and at $z/H=0.2$, where $Fr_{Hu}=0.021$}. The vertical black dash-dotted line in (a-d) corresponds to the CF--BF interface. Panels (e) and (f) show the radial average of $\langle u_{r}\rangle$ within $r/R < 0.5$ for anticyclones and cyclones, respectively.}
	\label{fig:vortex_motion}
\end{figure}

Furthermore, the vortex trajectories can be obtained by tracking the vortex center as the system evolves. Based on the trajectories, we then perform statistical analysis on the radial component for the vortex motion, which is directly connected to the centrifugal force. In figures \ref{fig:vortex_motion}(a)-(d), we present four profiles ($Fr= 0$, $0.02$, $0.12$, and $0.48$, respectively) of the vortex radial velocity component \chreplaced{$\langle u_r\rangle$}{ $\overline{u}_r$} for $\Gamma=1$ \chreplaced{(}{in}Set I\chadded{)}. Statistics are taken in the core flows, i.e., $r<r_{BF}$. In figure~\ref{fig:vortex_motion}(a), where the centrifugal force is absent, there is no significant deviation between cyclones and anticyclones. The negative \chreplaced{$\langle u_r\rangle$}{ $\overline{u}_r$} is attributed to the vortex emission from the boundary flow, as shown in figures \ref{fig:vortex_emit}. 
Clear separation between cyclones and anticyclones can be observed in figure~\ref{fig:vortex_motion}(c), corresponding to $Fr=0.12$. In this case, the radial velocity component for the anticyclones is all positive (towards the sidewall). As for the cyclones, negative motion (towards the axis) near the boundary flow is enhanced by strong centrifugal effects. But for small radius ($r/R\lesssim0.5)$, positive values close to those of anticyclone can be observed, which is attributed to the clustering effect. As $Fr$ further increases, the clustering effect reported in \citet{ding_nc_2021,ding_jfm_2023_vortex} becomes dominant, characterised by the positive velocity for both cyclones and anticyclones.

In figures \ref{fig:vortex_motion}(e) and \ref{fig:vortex_motion}(f), we plot the mean radial velocity of anticyclones and cyclones, respectively. 
To reduce the influence of vortex emission from the boundary flow, only data with $r/R\leq0.5$ are considered when computing the statistics of vortex trajectories. 
From figure~\ref{fig:vortex_motion}(e), a clear change in $\langle u_r\rangle$ from no preference ($\langle u_r\rangle\approx0$) to outward motion ($\langle u_r\rangle>0$) can be found once $Fr/Fr_{Hu}>1$. 
Such a result provides evidence for the interpretation of $Fr_{Hu}$---the emergence of bulk centrifugal effects. 
As for the cyclones shown in figure~\ref{fig:vortex_motion}(f), the change of $\langle u_r\rangle$ is not significant. This could be attributed to the competition between the centrifugal acceleration ($u_r<0$) and the clustering effect ($u_r>0$). 
We also find that the data for $\Gamma = 0.5$ are more scattered than those for the other values. 
The reason could be the lack of vortices in the statistical region $r/R<0.5$. 
Nevertheless, despite the additional effects, the radial vortex motion begins to be affected by the centrifugal force as $Fr$ exceeds $Fr_{Hu}$, which supports $Fr_{Hu}$ as the critical Froude number for the bulk flow.

%% file: sec_result_BL_onset.tex
\subsection{Centrifugal effects in the boundary layer}
\label{subsec:bl_onset}
We now turn to the onset of centrifugal effects in the \chadded{top and bottom thermal} boundary layer\chadded{s}, which is directly linked to the reduction of global heat transport.
We define $Fr_c^*$ as the critical Froude number at which $Nu$ begins to decrease.
Since global heat transport is governed by the thermal boundary layer in the classical regime, $Fr_c^*$ should then refer to the onset of centrifugal effects within the boundary layer.
In this section, we will provide evidence from both the boundary-layer thickness and local heat transport contributions to \chreplaced{support}{ verify} this picture.

\subsubsection{Quantitative definition of $Fr_c^*$ and boundary layer thickness}

\begin{figure}
	\centerline{\includegraphics[width=0.85\textwidth]{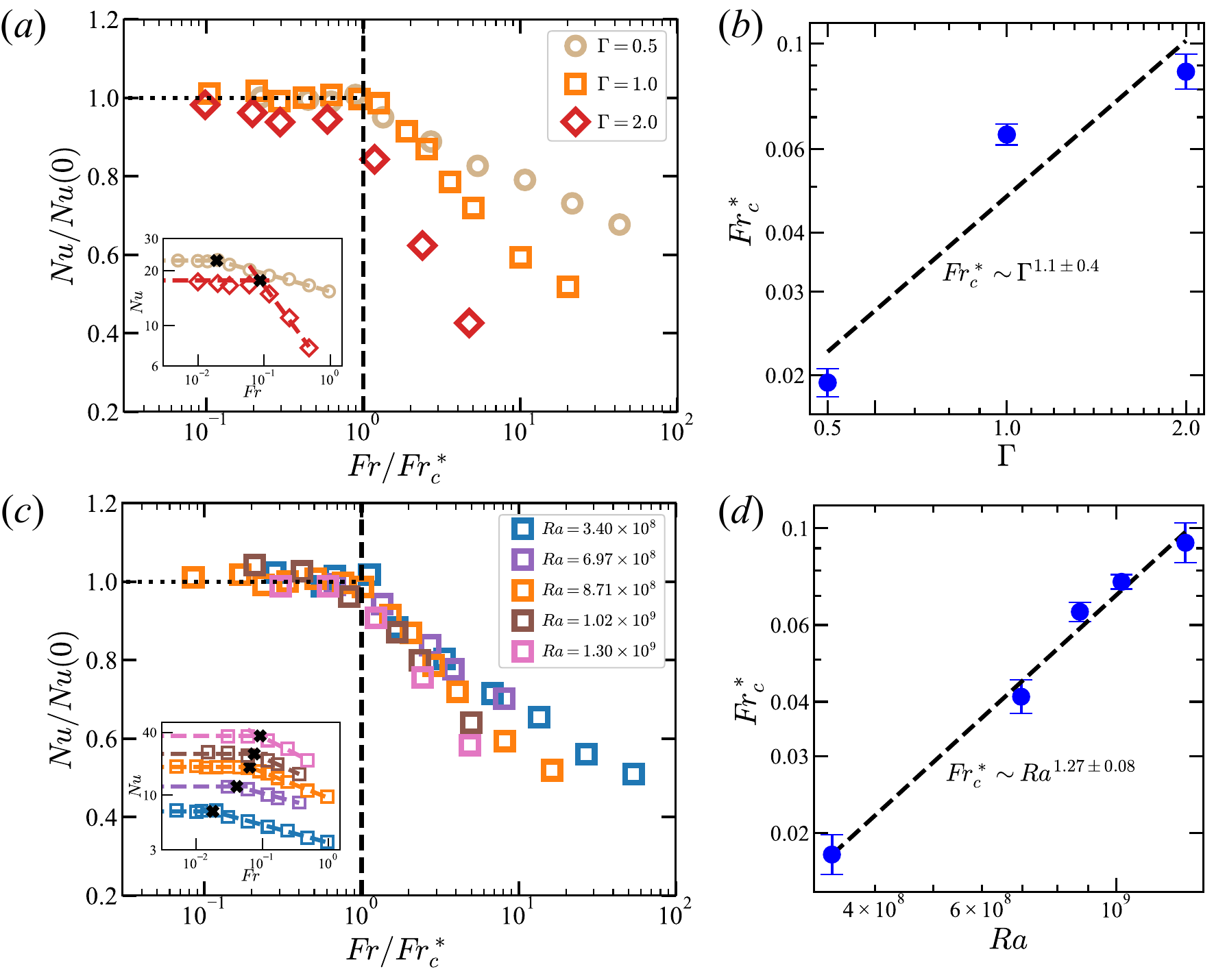}}
	\caption{(a,c) The plot of $Nu$ normalised by the non-centrifugal case $Nu(0)$ versus $Fr/Fr^*_c$ for Set I and II. The inset shows the fitting of $Fr^*_c$. The black dashed line indicates the critical Froude number $Fr_c^*$. The fitting result of $Fr^*_{c}$ versus (b) $\Gamma$ and (d) $Ra$. The corresponding power law behaviours are $Fr^*_c = 4.79 \times 10^{-2} \Gamma^{1.1\pm0.4}$ and  $Fr^*_c = 2.54 \times 10^{-13} Ra^{1.27\pm0.08}$, respectively.}
	\label{fig:nu_norm}
\end{figure}

In figures \ref{fig:nu_norm}(a) and \ref{fig:nu_norm}(c), we present the normalized heat transport efficiency $Nu/Nu(0)$ as a function of $Fr/Fr_c^*$ for Set I and II, respectively, where $Nu(0)$ are the heat transport of \chadded{the corresponding} cases \chreplaced{with $Fr=0$}{ without centrifugal effects}. 
The values of $Fr_c^{*}$ are determined from the intersection between $Nu(0)$ and linear fits of the $Nu-Fr$ curves in logarithmic coordinates, as indicated by the cross symbols in the insets. 
Although $Nu$ decreases for both sets when $Fr/Fr_c^*>1$, the behaviour differs between them. 
The case with larger $\Gamma$ in Set I declines more sharply, as shown in figure~\ref{fig:nu_norm}(a). 
As for Set II (cases with different $Ra$ in figure~\ref{fig:nu_norm}(c)), no significant $Ra$-dependence can be observed. 
\chadded{We note that,} for larger $\Gamma$, the increased proportion of core flow leads to a more rapid decay in the global heat transport. 
This is consistent with the local heat flux shown in figure~\ref{fig:decomp_nu}(a), where the convective heat flux in the core flow approaches zero under strong centrifugal forcing. 
In contrast, the boundary-flow contribution \chreplaced{decreases with a trend that appears to be levelling-off.}{ saturates}, as shown in figure~\ref{fig:decomp_nu}(b), indicating that the boundary flow is more robust than the core flow with increasing $Fr$. This explains the observed $\Gamma-$dependence of the global heat transport.

To summarise the dependence of $Fr_c^*$ on $\Gamma$ and $Ra$, we plot $Fr_c^*$ as a function of these parameters in figures \ref{fig:nu_norm}(b) and \ref{fig:nu_norm}(d). We find that $Fr_c^*\sim\Gamma^{1.1\pm 0.4}$, which is close to $Fr_{Horn} \sim \Gamma^{1}$. This agreement suggests a close relationship between the local onset of centrifugal effects and the global centrifugal-gravitational force balance. 
In addition, $Fr_c^*$ follows the scaling relation $Fr_c^*\sim Ra^{1.27\pm0.08}$ at fixed $\Gamma$. 

\begin{figure}
	\centerline{\includegraphics[width=0.9\textwidth]{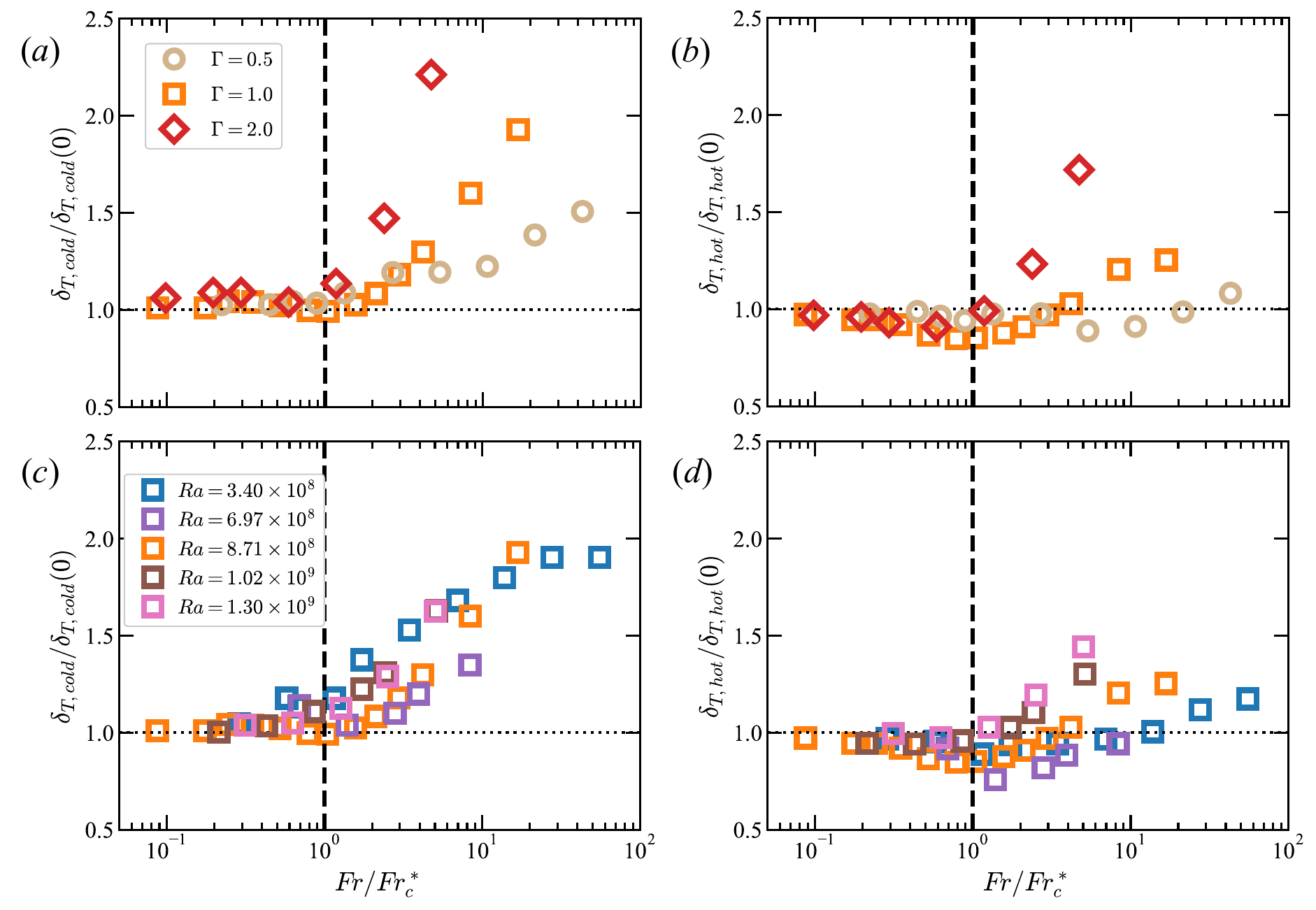}}
	\caption{Normalized thermal boundary layers thickness, $\delta_T/\delta_T(0)$, for the top (cold) and bottom (hot) planes as a function of $Fr/Fr_c^*$. Panels (a) and (c) correspond to the top boundary layer for Sets I and II, respectively, while (b) and (d) show the corresponding data for the bottom boundary layer. The thermal boundary layers thickness is obtained by the linear fitting method.}
	\label{fig:bl_thickness}
\end{figure} 

To further \chreplaced{justify}{ verify} our \chreplaced{interpretation}{ implication} for $Fr_c^*$, we examine the thickness of the thermal boundary layer. The thermal boundary layer thickness is obtained by the linear fitting method \textcolor{blue}{\citep{tilgner_pre_1993,lui_pre_1998}}. 
Figure~\ref{fig:bl_thickness} presents the normalised boundary layer thickness, $\delta_T/\delta_T(0)$ as a function of $Fr/Fr^*_c$ for both the top (cold) and bottom (hot) \chreplaced{plates}{ planes} in the left and right panels, respectively. 
The upper and lower panels correspond to Set I and II. 
In figures~\ref{fig:bl_thickness}(a) and \ref{fig:bl_thickness}(c), the boundary layer thickness for the top \chreplaced{plate}{ plane} exhibits a clear increase as $Fr/Fr_c^*$ exceeds unity. A similar trend is observed for the bottom boundary layer in figures~\ref{fig:bl_thickness}(b) and \ref{fig:bl_thickness}(d). 
However, an asymmetry between the top and bottom boundary layers can be observed by comparing figures~\ref{fig:bl_thickness}(a) with (b), \chreplaced{and}{ or} (c) with (d). 
The change in thermal boundary layer thickness for the hot plate is less sensitive to $Fr$ than that for the cold plate. 
Nevertheless, the change in boundary layer thickness further supports the argument that the boundary layers are directly influenced by centrifugal force when $Fr > Fr^*_c$.

\subsubsection{Local heat transport contribution}

\begin{figure}
	\centerline{\includegraphics[width=0.6\textwidth]{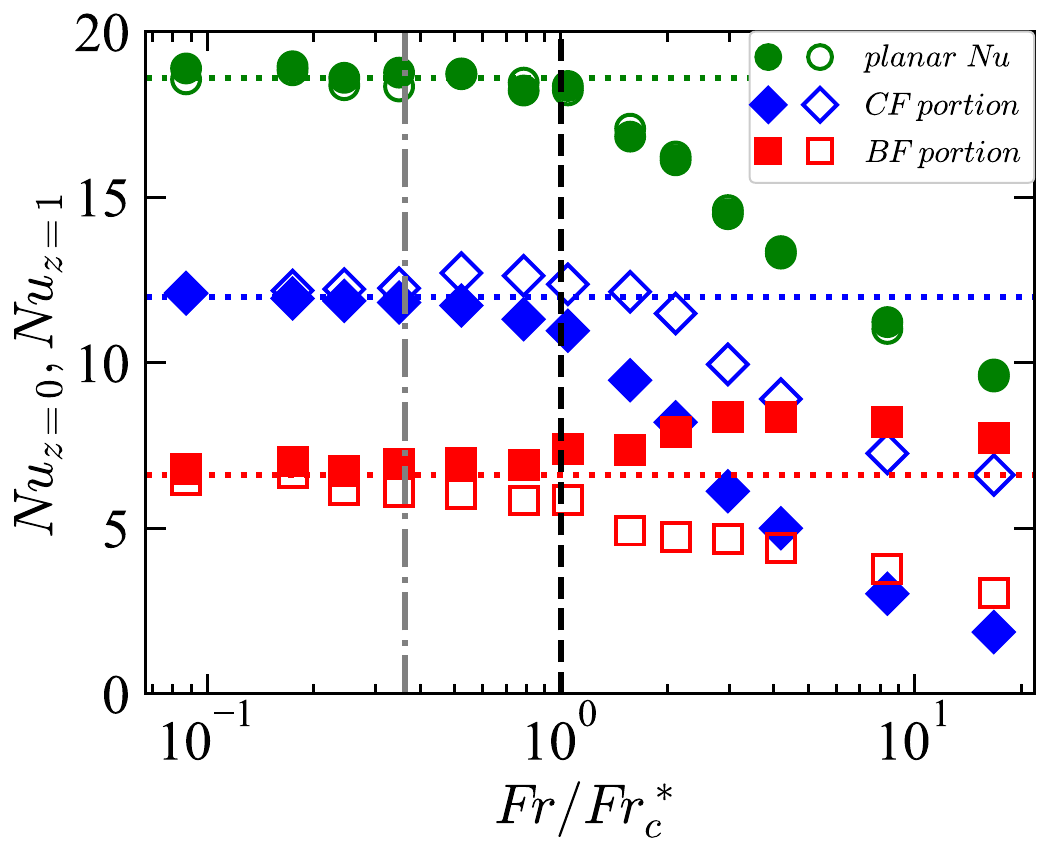}}
	\caption{Top (open symbols) and bottom (closed symbols) planar Nusselt number $Nu_{z=0}$ and $Nu_{z=1}$ (green circle), along with their portion of contributions from the core flow (blue diamond), boundary flow (red square), as a function of $Fr/Fr_c^*$ at $Ra=8.71\times10^{8}$ and $\Gamma=1$. The vertical grey dash-dotted line and the black dashed line indicate the critical Froude number $Fr_{Hu}$ and $Fr_c^*$, respectively.
    }
	\label{fig:plane_nu}
\end{figure}
To distinguish boundary-layer effects from bulk-induced redistribution, we analyse the local contributions to heat transport.
In figure~\ref{fig:plane_nu}, we present the planar Nusselt number $Nu_{z=0}$ and $Nu_{z=1}$ for $\Gamma=1$ in Set I. 
The subscripts $*_{z=0}$ and $*_{z=1}$ correspond to values averaged over the horizontal plates at $z=0$ (bottom plate) and $z=1$ (top plate), respectively.
Both the total planar Nusselt number and the contributions from the core and boundary flows are presented.
The planar Nusselt number, $Nu_{z=0}$ and $Nu_{z=1}$ (green circle), begins to decrease once $Fr>Fr_c^*$, which is, of course, consistent with those in figures \ref{fig:nu_norm}(a) and \ref{fig:nu_norm}(c), as the vertical heat flux should be statistically equal at each height. 
What we want to emphasise here are the portions of contributions from the core flow (CF) and the boundary flow (BF).  
The CF portion (blue diamonds) decreases sharply for $Fr/Fr_c^*>1$.  
In contrast, the BF portion (red squares) exhibits an asymmetric response: the top plane contribution (open squares) decreases, while the bottom one (closed squares) increases. 
This local enhancement in the bottom BF region can be understood as follows. 
The \chadded{flows in the} top (bottom) boundary layer is driven outwards (inward) by the centrifugal force. 
Thus, the boundary flow \chadded{(at the sidewall)} is overall downwelling when the centrifugal effects are strong. 
Such downwelling flow hits the bottom plate, leading to the enhancement in local heat transport. 
Slight deviations in both contributions can be observed for $Fr_{Hu}<Fr<Fr_c^*$.
\chadded{Such deviation can be attributed to radial vortex motion and the redistribution of heat flux induced by the bulk centrifugal effects.}
However, the unchanged global $Nu$ suggests that the centrifugal effects are not significant in the thermal boundary layer \chadded{as long as $Fr$ is less than $Fr^*_c$}. 
\chdeleted{Such deviation can be attributed to radial vortex motion and the redistribution of heat flux induced by the bulk centrifugal effects.}

For further \chreplaced{support the above physical picture}{ verification}, we examine the heat transfer through the boundary flow region $Q_{BF}(z)$ and the core flow region $Q_{CF}(z)$

\begin{equation}
        Q_{*}(z) \equiv \frac{\int_{S_{*}}\left(\sqrt{RaPr}u_z \theta -\partial \theta/\partial z\right)dA}{\pi R^2},
\end{equation}
where the subscripts $*$ denote the integral region (e.g. BF \chreplaced{or}{ and} CF). \chdeleted{This definition is similar to that in figure~\ref{fig:plane_nu} but considers the convective heat transfer in the bulk.}

\begin{figure}
	\centerline{\includegraphics[width=0.9\textwidth]{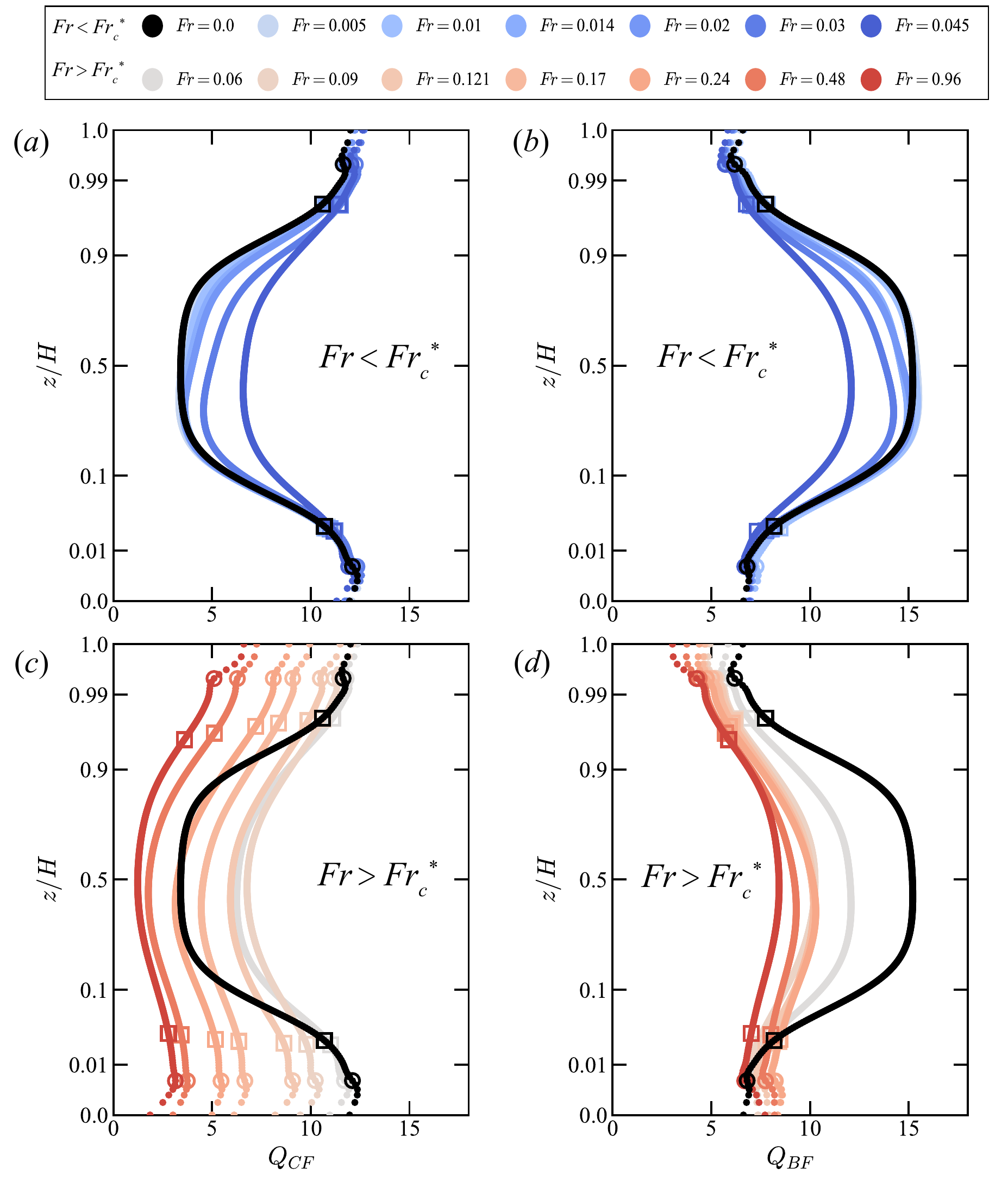}}
	\caption{\chreplaced{Contribution of (a,c) the core flow region $Q_{CF}$ and (b,d) the boundary flow region $Q_{BF}$, as a function of the normalized height $z/H$. For (a,b) $Fr<Fr_c^*$ and (c,b) $Fr>Fr_c^*$. All plots are for $Ra=8.71\times10^{8}$ and $\Gamma=1.0$, where $Fr_{Hu}\approx 0.021$ and $Fr_c^*\approx 0.057$.}{ Contribution of the boundary flow region $Q_{BF}$, as a function of the normalized height $z/H$ for (a) $Fr<Fr_c^*$ and (b) $Fr>Fr_c^*$.} The vertical coordinate is stretched using the mapping $z\to \text{sin}^2(\pi z/2)$ to highlight the top and bottom boundary layers. Large open circles and squares mark the Ekman and thermal boundary layer thicknesses, respectively. The Ekman boundary layer thickness is determined from the vertical position of the maximum of $u^{\text{rms}}_{\phi}$, while the thermal boundary layer thickness is obtained using the linear fitting method. \chdeleted{Insets in (a,b) show the contribution of the core flow region $Q_{CF}$}.
    }
	\label{fig:Q_ratio}
\end{figure}
\chreplaced{
As shown in figure~\ref{fig:Q_ratio}, the contributions of the core flow region $Q_{CF}(z)$ and the boundary flow region $Q_{BF}(z)$ are plotted as functions of height for different $Fr$. 
For $Fr<Fr_c^*$, figure~\ref{fig:Q_ratio} (a) and (b) present $Q_{CF}$ and $Q_{BF}$, respectively. 
In these cases, both $Q_{CF}(z)$ and $Q_{BF}(z)$ remain nearly unchanged within the thermal boundary layer (marked by a large open square), while a more pronounced variation is observed in the bulk region.
For $Fr_{Hu}<Fr<Fr_c^*$, radial shear motions are present in the bulk region; therefore, the deviation observed in figure~\ref{fig:plane_nu} (i.e., the difference between BF and CF contributions at the plates) is primarily governed by bulk dynamics rather than by direct centrifugal effects on the boundary layers. 
}
{
As shown in figure~\ref{fig:Q_ratio}, the contribution of the boundary flow region $Q_{BF}(z)$ is plotted as a function of height for different $Fr$. 
In figure~\ref{fig:Q_ratio}(a), where $Fr<Fr_c^*$, $Q_{BF}(z)$ remains nearly unchanged within the thermal boundary layer (marked by a large open square), while a more pronounced variation is observed in the bulk region. 
}In this regime, the global vertical heat transport, quantified by $Nu$, remains essentially unchanged. 
\chreplaced{
This indicates that the enhancement of $Q_{CF}$ in figure~\ref{fig:Q_ratio} (a) is associated with the radial heat transport discussed earlier, as evidenced by the reduction of $Q_{BF}(z)$ shown in figure~\ref{fig:Q_ratio} (b).
When $Fr>Fr_c^*$, figure~\ref{fig:Q_ratio} (c) and (d) show $Q_{CF}$ and $Q_{BF}$, respectively. 
A clear change emerges within the boundary layers, particularly in $Q_{CF}$ and, to a lesser extent, in $Q_{BF}$. 
}
{
This indicates that the reduction in $Q_{BF}(z)$ is associated with the radial heat transport discussed earlier, as evidenced by the inset data showing an enhancement of $Q_{CF}$ in the core flow.
When $Fr>Fr_c^*$, a clear change emerges within the boundary layers, particularly in CF contributions $Q_{CF}$ (inset of figure~\ref{fig:Q_ratio}(b)). 
}
\chdeleted{
For $Fr_{Hu}<Fr<Fr_c^*$, radial shear motions are present in the bulk region and thus the difference between the BF and CF contributions at the plates, shown in figure~\ref{fig:plane_nu}, is not a direct consequence of centrifugal effects acting on the boundary layers. }
Together with the reduction in global heat transport, these results demonstrate that the observed transition cannot be attributed solely to bulk redistribution, but instead arises from direct modifications of the top and bottom thermal boundary layers. 
Therefore, $Fr_c^*$ marks the onset of centrifugal effects within the boundary layers.

In summary, once $Fr$ exceeds $Fr_{Hu}$, the centrifugal effects redistribute heat transport within the bulk, while the global heat transport remains essentially unchanged. When $Fr>Fr_c^*$, the global heat transport is reduced. At this stage, both the boundary layer thickness and the local heat transport contributions are modified.

%% file: sec_conclusion.tex
\section{Conclusion}\label{sec:conclusion}
In summary, we \chreplaced{have conducted}{ conduct} a series of numerical simulations of rotating Rayleigh-Bénard convection (RRBC) in the presence of centrifugal force. 
We focus on the emergence of centrifugal effects in both the bulk and boundary layer, and their impact on the local and global heat transfer. 
Two sets of simulations are arranged to respectively explore the effects of $\Gamma$ and $Ra$. In each set, we vary $Fr$ and examine the system's heat transfer. 
The range of $Fr$ in the current study covers the critical Froude numbers proposed by \citet{hu_jfm_2022} and \citet{horn_prl_2018}.

The global impact of centrifugal effects is manifested as a reduction in global heat transport. 
We demonstrate in figure~\ref{fig:global_nu} that the critical Froude number for the decrease of $Nu$, namely $Fr_c^*$, cannot be clearly explained by either $Fr_{Hu}$ or $Fr_{Horn}$. 
To understand the mechanism, we decompose the vertical convective heat flux $\langle J_z\rangle$ into contributions from the core ($r<r_{BF}$) and the boundary flow ($r>r_{BF}$). 
Interestingly, although the global $Nu$ is unchanged for $Fr\gtrsim Fr_{Hu}$, the local $\langle J_z\rangle$ exhibit $Fr$-dependence once $Fr/Fr_{Hu}$ exceeds 1, as shown in figure~\ref{fig:decomp_nu}. 
This result can be understood as internal heat exchange between the core and boundary flows once the bulk centrifugal effects emerge. 
To verify this picture, we examine the evolution of vortex motion with increasing $Fr$, which serves as a key proxy of the bulk flow. 
In the absence of centrifugal force, vortices spontaneously form in the core. 
Meanwhile, numerous vortices are emitted from the boundary flow, producing a negative radial vortex velocity as shown in figure~\ref{fig:vortex_emit} and \ref{fig:vortex_motion}(a). 
When $Fr$ exceeds $Fr_{Hu}$ in figure~\ref{fig:vortex_motion}\chreplaced{(c)}{ (b)}-(d), the bulk centrifugal effects emerge, leading to the separation of cold and hot vortices. 
The mean radial vortex velocity of anticyclones (cold vortices) exhibits an apparent increase for $Fr/Fr_{Hu}>1$ in figure \ref{fig:vortex_motion}(e). 
For cyclones (hot vortices) in figure~\ref{fig:vortex_motion}(f), the inward centrifugal force is compensated by the outward clustering effect, as indicated in \citet{ding_nc_2021}. 
Thus, the mean radial vortex velocity of the cyclones (hot vortices) exhibits no clear dependence on $Fr$. 
Therefore, both the local heat flux and the radial vortex motion supports $Fr_{Hu}$ as the onset of bulk centrifugal effects in the RRBC system. 

Next, we focus on $Fr_c^*$, identified as the critical Froude number for the centrifugal effects within the thermal boundary layer. In figure~\ref{fig:nu_norm}, we present $\Gamma$- and $Ra$-dependence of $Fr_c^*$ in the current study, which can be described by the power laws $Fr^*_c = 4.79 \times 10^{-2} \Gamma^{1.1\pm0.4}$ for fixed $Ra=8.71\times 10^8$, and  $Fr^*_c = 2.54 \times 10^{-13} Ra^{1.27\pm0.08}$ for fixed $\Gamma=1$, respectively.  
However, it should be noted that the $Ra$-dependence may vary with $\Gamma$, and conversely, the $\Gamma$-dependence may also vary with $Ra$, therefore scaling relations may not hold universally across \chadded{the} parameter space.
For the first power law depending on $\Gamma$, the exponent 1.1 is close to $Fr_{Horn}\sim\Gamma^1$, suggesting a connection between the transition of global dominating mechanism and the local emergence of centrifugal effects within the thermal boundary layer. 
A theoretical explanation of these power laws is still required in future studies. 
Furthermore, the global heat transport is controlled by the thermal boundary layer in the current parameter range. 
Consequently, as the global Nusselt number declines, the thermal boundary layer should be affected by the centrifugal force. 
This is confirmed in figure~\ref{fig:bl_thickness}, where the normalized top and bottom thermal boundary layer thicknesses $\delta_T/\delta_T(0)$ clearly increase for $Fr>Fr_c^*$.
Locally, the bottom boundary layer thickness decreases. 
This could be attributed to the overall downwelling boundary flow since the top (bottom) boundary layer is driven outwards (inward) by the centrifugal effects. 
Consequently, the contribution from the boundary flow on the bottom planar heat transport is enhanced as shown in figure~\ref{fig:plane_nu}. 
Nevertheless, other local contributions to the planar heat flux decrease for $Fr>Fr_c^*$. 
Finally, figure~\ref{fig:Q_ratio} illustrates the contribution of the boundary flow region $Q_{BF}(z)$. 
The ratio within the boundary layer remains nearly constant when $Fr<Fr_c^*$, but changes significantly once $Fr>Fr_c^*$. Thus, both the boundary layer thickness and local contributions to heat transport support $Fr_c^*$ as the onset of the boundary layer centrifugal effects in the RRBC system.

\chadded{We would like to point out that the boundary flow structure persists and continues to dominate heat transport even at the largest Fr examined in the present study, as demonstrated in figures \ref{fig:decomp_nu} and \ref{fig:Q_ratio}. One may infer a breakdown of the boundary flow and reorganization of heat transport in this system for even larger $Fr$. This process would likely lead to a qualitatively different regime. How heat transport is subsequently re-established under such extreme centrifugal conditions remains unexplored and requires further investigation. The overall picture that emerges from this study is that the effect of centrifugal force is felt differently in the bulk and in the top and bottom boundary layers. It requires a larger centrifugal force to influence the boundary layer than is required for the bulk.}


%% file: sec_appendix.tex
\section{}
The data sets in this study are listed in table \ref{tab:data}.

\begin{longtable}{cccccccccccc}
\hline
\hline \\
$Ra$ & $Fr$ & $\Gamma$ & $Fr_{Hu}$ & $Fr_c^*$ & $Nu$ & $Nu_{z=1}$ & $Nu_{z=0}$ &
$\delta_{E,cold}$ & $\delta_{E,hot}$ & 
$\delta_{T,cold}$ & $\delta_{T,hot}$ \\ 
$\left(\times10^8\right)$ & & & & & & & & 
$\left(\times10^{-3}\right)$ & $\left(\times10^{-3}\right)$ 
& $\left(\times10^{-3}\right)$ & $\left(\times10^{-3}\right)$ \\
3.40  & 0      & 1.0 & 0.013 & 0.017 & 6.99  & 7.03  & 6.95  & 4.41 & 4.41 & 63.70  & 64.42 \\
3.40  & 0.005  & 1.0 & 0.013 & 0.017 & 7.15  & 7.09  & 7.09  & 4.46 & 4.45 & 66.65  & 62.42 \\
3.40  & 0.01   & 1.0 & 0.013 & 0.017 & 6.90  & 7.14  & 6.57  & 4.45 & 4.44 & 74.99  & 61.45 \\
3.40  & 0.012  & 1.0 & 0.013 & 0.017 & 7.09  & 6.53  & 7.23  & 4.48 & 4.44 & 72.28  & 59.31 \\
3.40  & 0.02   & 1.0 & 0.013 & 0.017 & 7.11  & 6.42  & 7.34  & 4.49 & 4.43 & 75.04  & 57.30 \\
3.40  & 0.03   & 1.0 & 0.013 & 0.017 & 6.18  & 6.11  & 6.42  & 4.45 & 4.43 & 87.61  & 60.52 \\
3.40  & 0.06   & 1.0 & 0.013 & 0.017 & 5.61  & 5.52  & 5.82  & 4.47 & 4.44 & 97.33  & 60.77 \\
3.40  & 0.121  & 1.0 & 0.013 & 0.017 & 4.99  & 5.02  & 5.08  & 4.51 & 4.45 & 107.12 & 62.37 \\
3.40  & 0.24   & 1.0 & 0.013 & 0.017 & 4.57  & 4.33  & 4.75  & 4.82 & 4.45 & 114.60 & 64.73 \\
3.40  & 0.48   & 1.0 & 0.013 & 0.017 & 3.91  & 3.97  & 4.00  & 4.50 & 4.45 & 121.25 & 72.06 \\
3.40  & 0.96   & 1.0 & 0.013 & 0.017 & 3.56  & 3.70  & 3.56  & 4.90 & 4.45 & 121.29 & 75.66 \\
\\
\hline \\
$Ra$ & $Fr$ & $\Gamma$ & $Fr_{Hu}$ & $Fr_c^*$ & $Nu$ & $Nu_{z=1}$ & $Nu_{z=0}$ &
$\delta_{E,cold}$ & $\delta_{E,hot}$ & 
$\delta_{T,cold}$ & $\delta_{T,hot}$ \\ 
$\left(\times10^8\right)$ & & & & & & & & 
$\left(\times10^{-3}\right)$ & $\left(\times10^{-3}\right)$ 
& $\left(\times10^{-3}\right)$ & $\left(\times10^{-3}\right)$ \\
6.97  & 0      & 1.0 & 0.018 & 0.043 & 12.12 & 12.37 & 12.27 & 4.27 & 4.42 & 39.25  & 40.02 \\
6.97  & 0.03   & 1.0 & 0.018 & 0.043 & 12.02 & 11.86 & 12.22 & 4.29 & 4.40 & 44.77  & 36.67 \\
6.97  & 0.06   & 1.0 & 0.018 & 0.043 & 11.43 & 11.79 & 11.71 & 4.29 & 4.41 & 40.75  & 30.36 \\
6.97  & 0.121  & 1.0 & 0.018 & 0.043 & 10.15 & 10.36 & 10.47 & 4.31 & 4.41 & 43.15  & 32.83 \\
6.97  & 0.17   & 1.0 & 0.018 & 0.043 & 9.40  & 9.47  & 9.53  & 4.33 & 4.40 & 47.06  & 35.37 \\
6.97  & 0.36   & 1.0 & 0.018 & 0.043 & 8.50  & 8.44  & 8.48  & 4.34 & 4.38 & 52.91  & 37.76 \\
\\
\hline \\
$Ra$ & $Fr$ & $\Gamma$ & $Fr_{Hu}$ & $Fr_c^*$ & $Nu$ & $Nu_{z=1}$ & $Nu_{z=0}$ &
$\delta_{E,cold}$ & $\delta_{E,hot}$ & 
$\delta_{T,cold}$ & $\delta_{T,hot}$ \\ 
$\left(\times10^8\right)$ & & & & & & & & 
$\left(\times10^{-3}\right)$ & $\left(\times10^{-3}\right)$ 
& $\left(\times10^{-3}\right)$ & $\left(\times10^{-3}\right)$ \\
8.71  & 0      & 0.5 & 0.021 & 0.022 & 22.69 & 23.05 & 23.12 & 4.48 & 4.56 & 21.90  & 22.77 \\
8.71  & 0.005  & 0.5 & 0.021 & 0.022 & 22.70 & 22.59 & 23.34 & 4.49 & 4.56 & 22.48  & 22.21 \\
8.71  & 0.01   & 0.5 & 0.021 & 0.022 & 22.59 & 22.69 & 22.80 & 4.50 & 4.56 & 22.48  & 22.50 \\
8.71  & 0.014  & 0.5 & 0.021 & 0.022 & 22.47 & 22.56 & 22.88 & 4.50 & 4.54 & 22.85  & 21.95 \\
8.71  & 0.02   & 0.5 & 0.021 & 0.022 & 22.85 & 23.11 & 23.42 & 4.51 & 4.55 & 22.66  & 21.44 \\
8.71  & 0.03   & 0.5 & 0.021 & 0.022 & 21.56 & 21.92 & 22.03 & 4.52 & 4.55 & 23.80  & 22.21 \\
8.71  & 0.0605 & 0.5 & 0.021 & 0.022 & 20.16 & 20.57 & 20.40 & 4.53 & 4.54 & 26.11  & 22.23 \\
8.71  & 0.12   & 0.5 & 0.021 & 0.022 & 18.76 & 18.63 & 19.11 & 4.53 & 4.50 & 26.14  & 20.23 \\
8.71  & 0.24   & 0.5 & 0.021 & 0.022 & 17.95 & 18.18 & 18.20 & 4.56 & 4.46 & 26.81  & 20.75 \\
8.71  & 0.48   & 0.5 & 0.021 & 0.022 & 16.59 & 16.15 & 17.10 & 4.56 & 4.49 & 30.35  & 22.44 \\
8.71  & 0.96   & 0.5 & 0.021 & 0.022 & 15.36 & 15.35 & 15.45 & 4.53 & 4.50 & 32.99  & 24.64 \\
\\
\hline \\
$Ra$ & $Fr$ & $\Gamma$ & $Fr_{Hu}$ & $Fr_c^*$ & $Nu$ & $Nu_{z=1}$ & $Nu_{z=0}$ &
$\delta_{E,cold}$ & $\delta_{E,hot}$ & 
$\delta_{T,cold}$ & $\delta_{T,hot}$ \\ 
$\left(\times10^8\right)$ & & & & & & & & 
$\left(\times10^{-3}\right)$ & $\left(\times10^{-3}\right)$ 
& $\left(\times10^{-3}\right)$ & $\left(\times10^{-3}\right)$ \\
8.71  & 0      & 1.0 & 0.021 & 0.057 & 18.59 & 18.61 & 18.73 & 4.52 & 4.59 & 24.88  & 25.24 \\
8.71  & 0.005  & 1.0 & 0.021 & 0.057 & 18.78 & 18.56 & 19.03 & 4.52 & 4.59 & 25.14  & 24.54 \\
8.71  & 0.01   & 1.0 & 0.021 & 0.057 & 18.92 & 18.83 & 19.10 & 4.53 & 4.60 & 25.12  & 23.89 \\
8.71  & 0.014  & 1.0 & 0.021 & 0.057 & 18.43 & 18.38 & 18.74 & 4.54 & 4.59 & 25.96  & 23.85 \\
8.71  & 0.02   & 1.0 & 0.021 & 0.057 & 18.60 & 18.34 & 18.90 & 4.54 & 4.59 & 25.87  & 23.21 \\
8.71  & 0.03   & 1.0 & 0.021 & 0.057 & 18.73 & 18.74 & 18.87 & 4.54 & 4.59 & 25.44  & 21.89 \\
8.71  & 0.045  & 1.0 & 0.021 & 0.057 & 18.25 & 18.52 & 18.64 & 4.55 & 4.61 & 24.80  & 21.40 \\
8.71  & 0.06   & 1.0 & 0.021 & 0.057 & 18.32 & 18.22 & 18.53 & 4.55 & 4.60 & 24.70  & 21.53 \\
8.71  & 0.09   & 1.0 & 0.021 & 0.057 & 16.99 & 17.07 & 17.00 & 4.53 & 4.59 & 25.51  & 22.12 \\
8.71  & 0.121  & 1.0 & 0.021 & 0.057 & 16.15 & 16.23 & 16.28 & 4.53 & 4.59 & 26.97  & 22.93 \\
8.71  & 0.17   & 1.0 & 0.021 & 0.057 & 14.62 & 14.61 & 14.64 & 4.54 & 4.53 & 29.37  & 24.49 \\
8.71  & 0.24   & 1.0 & 0.021 & 0.057 & 13.38 & 13.29 & 13.52 & 4.55 & 4.51 & 32.29  & 25.97 \\
8.71  & 0.48   & 1.0 & 0.021 & 0.057 & 11.05 & 11.02 & 11.36 & 4.58 & 4.51 & 39.81  & 30.40 \\
8.71  & 0.96   & 1.0 & 0.021 & 0.057 & 9.65  & 9.64  & 9.71  & 4.58 & 4.54 & 47.98  & 31.64 \\
\\
\hline \\
$Ra$ & $Fr$ & $\Gamma$ & $Fr_{Hu}$ & $Fr_c^*$ & $Nu$ & $Nu_{z=1}$ & $Nu_{z=0}$ &
$\delta_{E,cold}$ & $\delta_{E,hot}$ & 
$\delta_{T,cold}$ & $\delta_{T,hot}$ \\ 
$\left(\times10^8\right)$ & & & & & & & & 
$\left(\times10^{-3}\right)$ & $\left(\times10^{-3}\right)$ 
& $\left(\times10^{-3}\right)$ & $\left(\times10^{-3}\right)$ \\
8.71  & 0      & 2.0 & 0.021 & 0.103 & 17.66 & 17.77 & 17.30 & 4.34 & 4.46 & 23.92  & 23.93 \\
8.71  & 0.01   & 2.0 & 0.021 & 0.103 & 17.35 & 17.14 & 17.58 & 4.34 & 4.45 & 25.38  & 23.17 \\
8.71  & 0.02   & 2.0 & 0.021 & 0.103 & 16.99 & 16.93 & 17.31 & 4.34 & 4.44 & 26.05  & 22.98 \\
8.71  & 0.03   & 2.0 & 0.021 & 0.103 & 16.57 & 16.56 & 17.26 & 4.34 & 4.45 & 26.06  & 22.27 \\
8.71  & 0.06   & 2.0 & 0.021 & 0.103 & 16.69 & 16.43 & 17.24 & 4.33 & 4.46 & 24.88  & 21.75 \\
8.71  & 0.12   & 2.0 & 0.021 & 0.103 & 14.90 & 14.55 & 15.34 & 4.33 & 4.44 & 27.16  & 23.72 \\
8.71  & 0.242  & 2.0 & 0.021 & 0.103 & 11.02 & 10.75 & 11.17 & 4.36 & 4.40 & 35.20  & 29.49 \\
8.71  & 0.48   & 2.0 & 0.021 & 0.103 & 7.53  & 7.46  & 8.31  & 4.40 & 4.39 & 52.89  & 41.10 \\
\\
\hline \\
$Ra$ & $Fr$ & $\Gamma$ & $Fr_{Hu}$ & $Fr_c^*$ & $Nu$ & $Nu_{z=1}$ & $Nu_{z=0}$ &
$\delta_{E,cold}$ & $\delta_{E,hot}$ & 
$\delta_{T,cold}$ & $\delta_{T,hot}$ \\ 
$\left(\times10^8\right)$ & & & & & & & & 
$\left(\times10^{-3}\right)$ & $\left(\times10^{-3}\right)$ 
& $\left(\times10^{-3}\right)$ & $\left(\times10^{-3}\right)$ \\
10.20 & 0      & 1.0 & 0.023 & 0.070 & 24.85 & 25.12 & 25.30 & 4.50 & 4.59 & 16.66  & 17.43 \\
10.20 & 0.015  & 1.0 & 0.023 & 0.070 & 25.88 & 25.37 & 26.27 & 4.53 & 4.61 & 16.69  & 16.48 \\
10.20 & 0.03   & 1.0 & 0.023 & 0.070 & 25.49 & 25.20 & 25.94 & 4.53 & 4.61 & 17.26  & 16.41 \\
10.20 & 0.06   & 1.0 & 0.023 & 0.070 & 23.90 & 23.17 & 24.95 & 4.51 & 4.58 & 18.48  & 16.69 \\
10.20 & 0.121  & 1.0 & 0.023 & 0.070 & 21.63 & 21.29 & 22.34 & 4.50 & 4.57 & 20.39  & 17.89 \\
10.20 & 0.17   & 1.0 & 0.023 & 0.070 & 19.86 & 19.92 & 20.10 & 4.51 & 4.55 & 21.81  & 19.23 \\
10.20 & 0.36   & 1.0 & 0.023 & 0.070 & 15.91 & 15.82 & 16.02 & 4.56 & 4.52 & 27.11  & 22.70 \\
\\
\hline \\
$Ra$ & $Fr$ & $\Gamma$ & $Fr_{Hu}$ & $Fr_c^*$ & $Nu$ & $Nu_{z=1}$ & $Nu_{z=0}$ &
$\delta_{E,cold}$ & $\delta_{E,hot}$ & 
$\delta_{T,cold}$ & $\delta_{T,hot}$ \\ 
$\left(\times10^8\right)$ & & & & & & & & 
$\left(\times10^{-3}\right)$ & $\left(\times10^{-3}\right)$ 
& $\left(\times10^{-3}\right)$ & $\left(\times10^{-3}\right)$ \\
13.00 & 0      & 1.0 & 0.026 & 0.095 & 37.00 & 38.60 & 36.69 & 4.29 & 4.44 & 12.22  & 11.75 \\
13.00 & 0.03   & 1.0 & 0.026 & 0.095 & 36.53 & 36.52 & 37.05 & 4.29 & 4.43 & 12.68  & 11.70 \\
13.00 & 0.06   & 1.0 & 0.026 & 0.095 & 36.54 & 36.12 & 37.15 & 4.29 & 4.43 & 12.82  & 11.46 \\
13.00 & 0.121  & 1.0 & 0.026 & 0.095 & 33.56 & 33.55 & 33.99 & 4.31 & 4.42 & 13.75  & 12.10 \\
13.00 & 0.24   & 1.0 & 0.026 & 0.095 & 27.94 & 28.18 & 28.29 & 4.31 & 4.39 & 15.76  & 14.03 \\
13.00 & 0.48   & 1.0 & 0.026 & 0.095 & 21.59 & 21.56 & 21.90 & 4.35 & 4.37 & 19.87  & 16.96 \\ 
\\
\hline
\hline
\caption{Simulation parameters, corresponding critical Froude number, Nusselt number and the boundary layer thickness.} 
\label{tab:data}
\end{longtable}